\documentclass{article}

\usepackage[square,numbers]{natbib}
\usepackage[utf8]{inputenc} % allow utf-8 input
\usepackage{hyperref}       % hyperlinks
\usepackage{url}            % simple URL typesetting
\usepackage{booktabs}       % professional-quality tables
\usepackage{amsfonts}       % blackboard math symbols
\usepackage{microtype}      % microtypography
\usepackage{graphicx}       % To insert images
\usepackage{xcolor}
\usepackage{multirow}
\usepackage{booktabs}
\usepackage[margin=0.8in]{geometry}
\usepackage{authblk}

\title{Simulation-supervised deep learning for analysing organelles states and behaviour in living cells}

\author[1]{Arif Ahmed Sekh}
\author[1]{Ida S. Opstad}
\author[2]{Rohit Agarwal}
\author[3]{{\AA}sa Birna Birgisdottir}
\author[3]{Truls Myrmel}
\author[1]{Balpreet Singh Ahluwalia}
\author[1]{Krishna Agarwal}
\author[4]{Dilip K. Prasad}
\affil[1]{Department of Physics and Technology,UiT The Arctic University of Norway}
\affil[2]{Indian Institute of Technology (ISM),Dhanbad, India}
\affil[3]{Department of Clinical Medicine, UiT The Arctic University of Norway}
\affil[4]{Department of Computer Science, UiT The Arctic University of Norway }
\date{}
\begin{document}
\maketitle
\begin{abstract}

  In many real-world scientific problems, generating ground truth (GT) for supervised learning is almost impossible. The causes include limitations imposed by scientific instrument, physical phenomenon itself, or the complexity of modeling. Performing artificial intelligence (AI) tasks such as segmentation, tracking, and analytics of small sub-cellular structures such as mitochondria in microscopy videos of living cells is a prime example. The 3D blurring function of microscope, digital resolution from pixel size, optical resolution due to the character of light, noise characteristics, and complex 3D deformable shapes of mitochondria, all contribute to making this problem GT hard. Manual segmentation of 100s of mitochondria across 1000s of frames and then across many such videos is not only herculean but also physically inaccurate because of the instrument and phenomena imposed limitations. Unsupervised learning produces less than optimal results and accuracy is important if inferences relevant to therapy are to be derived. In order to solve this unsurmountable problem, we bring modeling and deep learning to a nexus. We show that accurate physics based modeling of microscopy data including all its limitations can be the solution for generating simulated training datasets for supervised learning. We show here that our simulation-supervised segmentation approach is a great enabler for studying mitochondrial states and behaviour in heart muscle cells, where mitochondria have a significant role to play in the health of the cells. We report unprecedented mean IoU score of 91\% for binary segmentation (19\% better than the best performing unsupervised approach) of mitochondria in actual microscopy videos of living cells. We further demonstrate the possibility of performing multi-class classification, tracking, and morphology associated analytics at the scale of individual mitochondrion.

\end{abstract}
\section{Introduction}
\label{sec1}
Computer vision (CV) is rapidly catching up in the field of microscopy, allowing fast and accurate segmentation and tracking of living cells \cite{chang2019kunet,zhou2018unet++,van2016deep,sadanandan2017automated,aydin2017cnn,majurski2019cell,sekh2020learning}. While cell-level studies are becoming routine, CV of sub-cellular structures is far more challenging in general because the structures are often in the order of $100-1000$ nm, while the resulting pixel size in microscopes may be comparable at $80 - 120$ nm and the optical resolution limit of advanced live cell compatible microscopes is in the range $200 - 300$ nm. Therefore, the details of the structures are often lost and hence segmenting is a significant challenge. Further, since these structures are visualized via labeling with fluorescent molecules, emitting only a few $100$ photons every $10-100$ ms (a typical exposure time in the case of dynamic sub-cellular structures), the signal to noise ratio is quite poor (generally in the range $2 - 4$), which definitely does not help the situation. Here, we consider one type of sub-cellular structures called mitochondria which is extremely challenging for CV when studied in living cells and of significant interest to life sciences \cite{vyas2016mitochondria,mills2017mitochondria,wong2018mitochondria,zamponi2018mitochondrial}.

\begin{figure}[!htb]
\centering
\includegraphics[width=\linewidth]{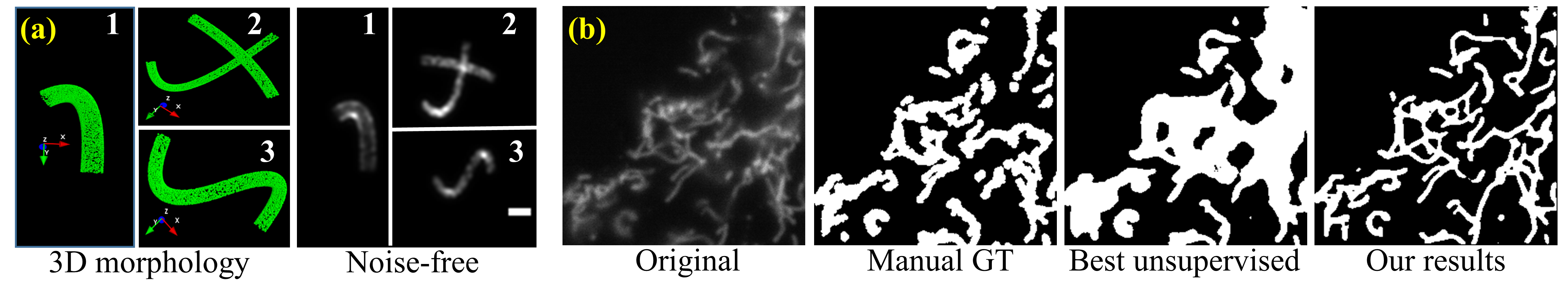}
\caption{(a) Challenges for manual segmentation
 (scale bar: 1 $\mu m$). (b) Our segmentation outperforms manual segmentation and best unsupervised approach.}
\label{fig:gt}
\end{figure}

The main challenges for CV in the case of mitochondria, are extreme morphological variability, three-dimensional distribution and dynamics, together with frequent structural morphing, merging and splitting. They are highly deformable 3D tube like structures with diameters in the range $100-800$ nm and lengths in the range $100$ nm - $5 \mu$m. Their 2D image formed by a microscope significantly differs with their 3D orientation and deformation, whether they are individually present or are in a network. Even if we ignore variability introduced by fluorescent labeling and optics of the microscope, two significant aspects cannot be ignored. These are the limitations introduced due to resolution and out-of-focus light. Both of them are related to the optical blurring introduced by the microscope, which is represented by the three-dimensional point spread function (3D-PSF). The 3D-PSF maps the photons emitted from a particular point in the 3D sample space to the 2D image plane including the effects of the optical lenses and the scattering phenomenon of light. It causes even a single molecule ($\sim 1$ nm) to appear as a comparatively large spot ($\sim 250$ nm diameter, referred to as optical resolution limit) in the image and inhibits resolving two structures closer than the spot size. Therefore a thin mitochondrion appears quite similar to a thick mitochondrion (see Figure \ref{fig:gt}(a), 1 and 3). Further, the image of a molecule that is away from the focal plane is significantly blurrier than the one in the focal plane. As a consequence, portions of mitochondria that are out of focus appear distorted and of low intensity, which either results in incorrect segmentation for those portions or their complete disappearance from segmentation. A third problem is witnessed in the case of mitochondria with overlapping (i.e. they are spaced closely in the axial direction compared to the objective lens' z-sectioning ability), where the local intensity in the region of overlap may be significantly larger than the intensity from individual mitochondria, posing difficulty in segmentation of the low intensity portions. Therefore, neither manual nor unsupervised segmentations of these structures are accurate enough. An example is illustrated in Figure \ref{fig:gt}(b).

Nonetheless, an expert may use software such as CellProfiler, ImageJ, Imaris, etc. to create reasonable ground truth. This comes in the scope of manual or semi-supervised segmentation. In fact, tools for automated analysis of mitochondrial shape and morphology have been proposed ~\cite{nikolaisen2014automated,lihavainen2012mytoe,ekanayake2015imaging}, but are deemed heavily dependent on thresholds employed by experts and often resulting into erroneous segmentation. Nonetheless, this is a feasible approach if there are only a few image frames per living cell and only a few cells to investigate \cite{mirzapoiazova2019monitoring}. However, for long duration videos with hundreds to thousand frames of a large number of cells, segmentation of every mitochondrion is a herculean task and a big impediment in conducting studies like tracking or morphological data analytics. Indeed, deep learning can fill the gap if large supervised training datasets can be formed and ground truth (GT) annotations can be derived for them. However, the problem of manual segmentation being both inaccurate and tedious stands as the road block for generating GT. Even for testing purposes, or performance quantification, we have no solution that is sufficiently accurate. This challenge demands unconventional solutions.

We solve this problem using a new concept of simulation-supervised training of deep learning networks. We generate supervised training datasets by simulating images of mitochondria as realistically as possible and using the morphology of the mitochondria itself, rather than the microscopy image, to create the ground truth. This is a significant challenge since the entire physics from mitochondria geometry, fluorescence labeling, photon emission, to the microscope's 3D PSF and noise characteristics have to be simulated. Such simulation engine and physics based ground truth is not only the first of its kind, it also generates a significantly better segmentation than the expert-generated manual ground truth (see Figure \ref{fig:gt}(b)).

\section{Related work}
The CV community uses object segmentation in image and video analysis and the microscopy community also employs segmentation of cells or sub-cellular structures for microscopy image and video analysis. There are two modalities of segmentation in CV, namely binary and semantic segmentation. Binary segmentation generates a binary mask, whereas semantic segmentation generates a different color mask of each class (multi-class segmentation). They are solved using supervised and unsupervised methods~\cite{kanezaki2018unsupervised}. The bottleneck of obtaining GT is absent in unsupervised methods~\cite{ragothaman2016unsupervised}. However, their poor performance in many complex cases implies that supervised methods, such as convolutional neural network (CNN)~\cite{kim2019cnn,deng2017cnn}, U-Net~\cite{ronneberger2015u} and generative models ~\cite{zhang2018clothingout,chen2018attention} are more popular.
It is therefore not surprising that CNN and other deep learning techniques have been explored for microscopy data as well. U-Net~\cite{ronneberger2015u} and deep CNN~\cite{sadanandan2017automated} have been successfully applied on cell segmentation since generating supervised training datasets is possible. Variation of semantic segmentation~\cite{noh2015learning}, Pix2Pix GAN~\cite{tsuda2019cell}, and Mask R-CNN method~\cite{he2017mask} have also been applied on multi-channel (i.e. multiple types of fluorescent markers highlighting different structures), multi-class segmentation problems. Nonetheless, manual supervised or unsupervised generation of ground truth for microscopy data is still tedious and sensitive to variability of cell appearances across microscopes and subjectivity of manual annotation. Hence, unsupervised methods such as Otsu thresholding ~\cite{xu2011characteristic} and different ImageJ based semi-supervised methods are still popular for several segmentation tasks.

Although the concept of generating training datasets using simulators for CV is quite new, there are some recent precedents in both microscopy and other GT hard problems. Datasets for tracking unmanned aerial vehicles \cite{mueller2016benchmark}, climate \cite{racah2017extremeweather}, object interaction and physical event prediction \cite{wu2015galileo,watters2017visual}, image understanding~\cite{zhang2019raven}, and crowd behaviour \cite{cheung2016lcrowdv} were synthesized using physics or other behaviour emulating approaches. Within microscopy, we are aware of a microscopy image simulator ~\cite{svoboda2017multimodal}, synthetic dataset for cells \footnote{https://data.broadinstitute.org/bbbc/image\_sets.html}, as well as a dataset for sub-cellular spherical structures called vesicles \cite{chenouard2014objective}. We note that the simulation of microscopy data of sub-cellular structures is far more challenging compared to other CV domain problems due to the finer and more complex nature of microscopy data and the need for emulating multiple physical and chemical phenomena for a single simulator.

Here, we present our contributions to these bodies of work. The most fundamental contribution of our work is to demonstrate that the concept of simulation-supervised deep learning can indeed solve the challenges mentioned in section \ref{sec1}. To that end, we have also contributed the following:

\textbf{(A) Generating the right simulator (emulating microscopy data with sufficient accuracy and details):} Simulating microscopy images of mitochondria is a challenging task. We model all details from the 3D deformable tubular structure of mitochondria, the fluorescent labels and their photokinetics, microscope properties and the noise in microscopy data. This is the first time such an extensive simulator has been designed for mitochondria and its segmentation.

\textbf{(B) Generating the GT:} We use the projection of the actual 3D morphology of the mitochondria onto the microscope's image plane to generate a physical GT. This solves the challenge of inaccurate ground truth during manual segmentation due to limitations of the microscope. This ground truth is successfully used in various supervised deep learning methods for binary and multi-class segmentation.

\textbf{(C) Creating large custom designed datasets for benchmarking:} The proposed physics based simulation and GT generation methods enable creation of large volume datasets with varying shape, position, morphology, and noise. It is also straight forward to generate such datasets for different microscope parameters. The datasets can be used to benchmark binary and multi-class segmentation models without the need for manual annotation. In fact, the quality of the segmentation achieved by our approach is far better than the manually generated GT for real images, as shown in Figure \ref{fig:gt}.

\textbf{(D) Performing a variety of advanced AI tasks from segmentation of mitochondria:} We demonstrate two applications, namely deriving morphological data analytics of mitochondria and event detection by the tracking of mitochondria. This becomes possible due to the reasonably accurate ($\sim 90\%$) segmentation. We therefore open up new possibilities for mitochondria analysis in living cells that will enable new insight into different aspects of fundamental cell biology.

\begin{figure}[htb]
\includegraphics[width=\linewidth]{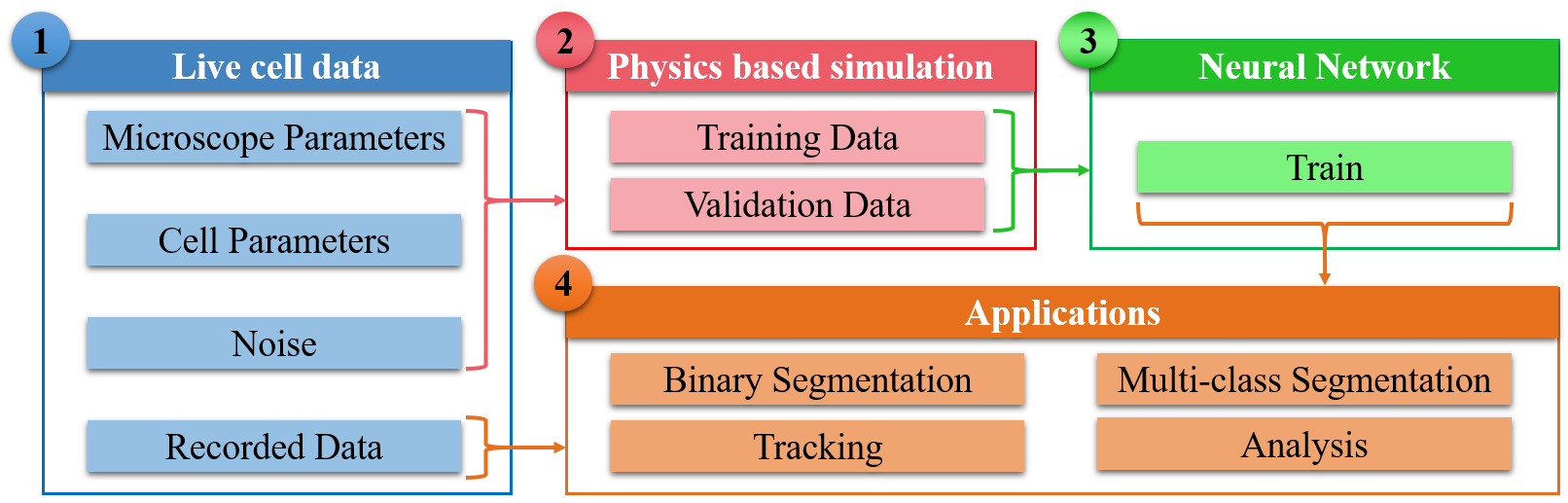}
\caption{Modules of the proposed method.}
\label{fig:method}
\includegraphics[width=\linewidth]{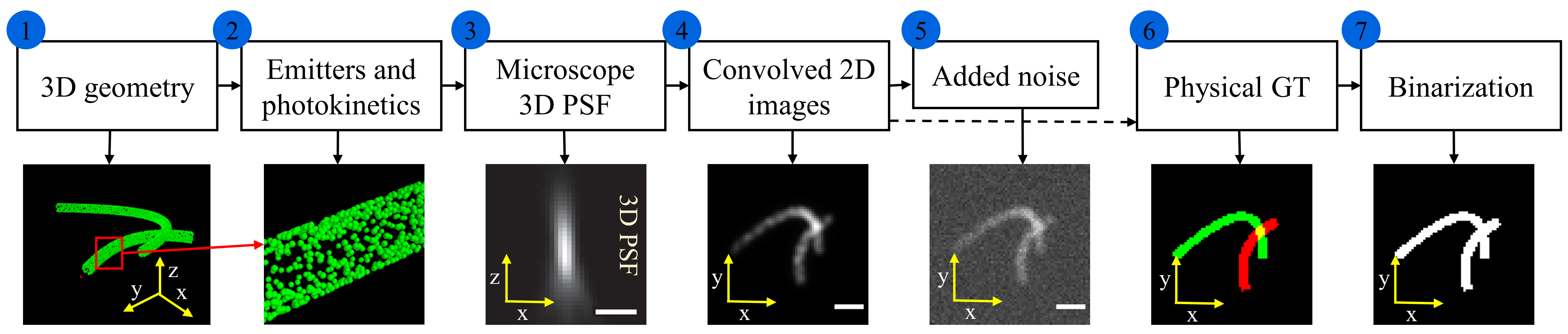}
\caption{Our simulation approach for modeling mitochondria. Scale bar: 1 $\mu m$.}
\label{fig:simulation}
\end{figure}

\section{Method}
Our methodology, shown in Figure \ref{fig:method}, consists
of four modules: (1) a live cell microscopy dataset, (2) a simulator for generating training data, (3) neural network to be trained for segmentation, and (4) a set of CV applications for analysis. We discuss modules 2-4 below. % and present details of our live cell dataset in the experiments section.

\subsection{Physics based simulation}
Our simulation flowchart is shown in Figure~\ref{fig:simulation}. We describe here briefly each of the steps in this seven-step process below, with more details present in the supplementary.
\\
\textcolor{white}{space}\textbf{1. Mitochondria geometry:} The process begins with a skeleton in 2D. $n$ number of control points ($3 \leq n \leq 5$) chosen as the ``knots", such that the boundary of the points is restricted by user-defined parameters. The skeleton of the mitochondria is defined by a 2D cubic spline of degree 3 passing through the control points. Next, the 2D cubic spline is converted into a 3D by a uniform distribution of $z$. $z$ is randomly taken from a user-defined parameters (z low, z high). The final geometry of the mitochondria is defined by convolving a 3D cylinder of radius ($r$) around the curve.
\\
\textcolor{white}{space}\textbf{2. Emitters and their photokinetics:} Fluorescent molecules, referred to as emitters, are randomly placed on the surface of the cylinder in a uniformly distributed manner such that the density of emitters is $\sim1000$ emitters/$\mu$m$^2$. For each emitter, the number of emitted photons are simulated using the photokinetics model~\cite{girsault2016sofi} includes blinking and non-radiative energy dissipation of the emitters~\cite{deschout2016complementarity}.
\\
\textcolor{white}{space}\textbf{3-4. Microscope's PSF and forming convolved 2D images:} We compute the 3D point spread function (PSF) of the microscope using a modern implementation \cite{li2017fast} of the Gibson-Lanni Model \cite{gibson1989diffraction}. We note that the PSF is specific to the type of microscope, its optical parameters, and the fluorescent dye used for labeling the mitochondria. We have modeled here an epi-fluorescence microscope, but the concept is generalizable for other microscopes as well. The 3D PSF is convolved over all the emitters to form the 2D image.
\\
\textcolor{white}{space}\textbf{5. Introducing noise:} Here, we note that since the exposure time is typically $10-100$ ms and the average number of photons emitted by an emitter is in the range $100-1000$ photons during these exposure time, the intensity in the image region is also in the order $100-2000$, which is significantly smaller than the bit depth of cameras. Therefore, the microscopy images are significantly more prone to noise than the conventional images for which most previous computer vision techniques were developed. Conventionally, modeling only shot noise is considered sufficient for images. However, because of low intensities, the camera's dark current noise is significant and comparable to shot noise. Therefore, we emulate both the dark current and shot noise in our model such that the signal to noise ratio of the simulated images matches the experimental images closely. More details about the observed and simulated signal to noise ratio are provided in the supplementary.
\\
\textcolor{white}{space}\textbf{6-7. Physical GT and binarization:} This is an important feature of our method and is therefore elaborated in the section \ref{sec:GT}.

\begin{figure}
\centering
    \begin{minipage}{0.445\textwidth}
    \includegraphics[width=\linewidth]{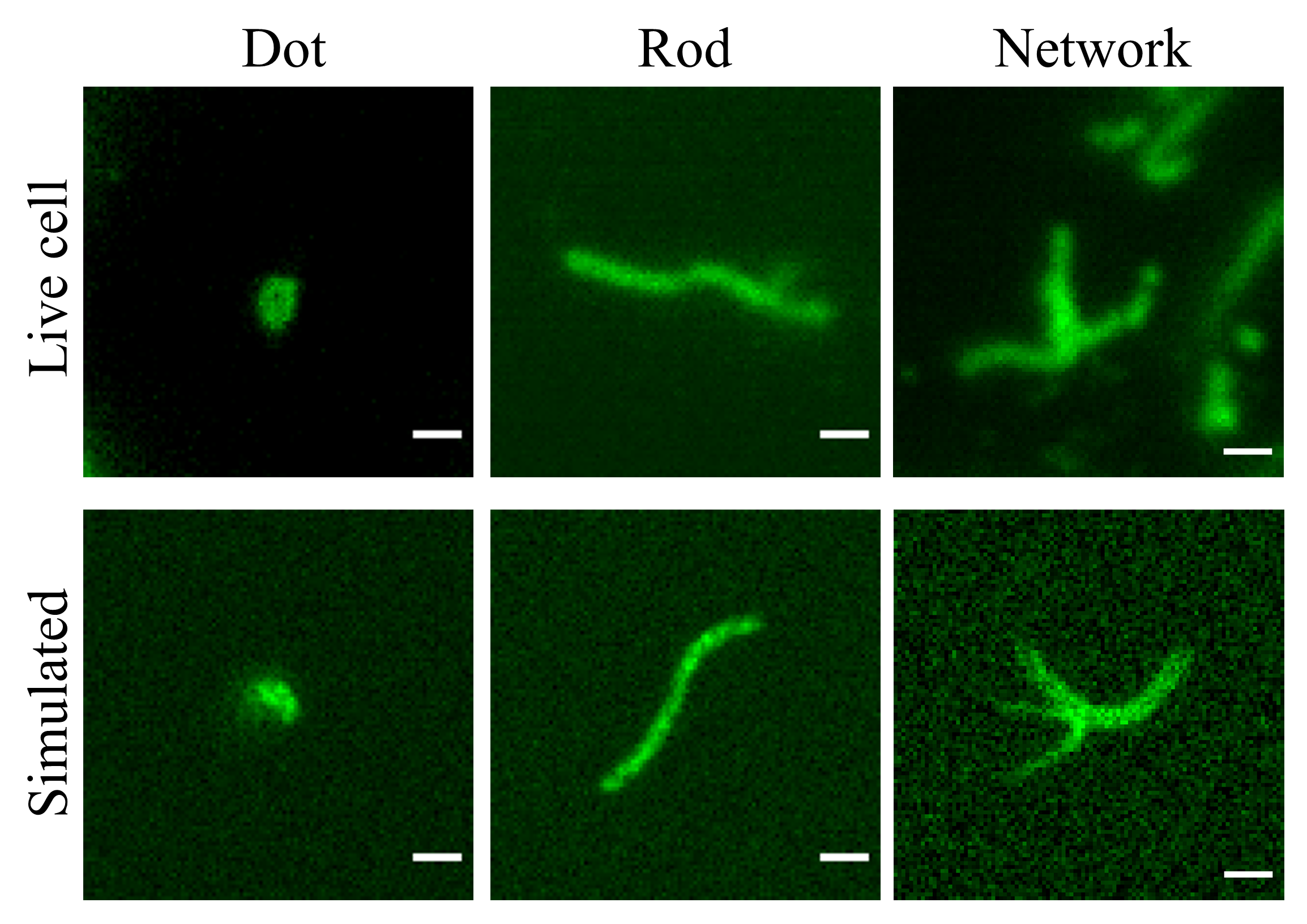}
    \caption{Three types of mitochondria shapes observed in live-cell images and simulated here (scale bar: 1 $\mu m$).}\label{fig:dataset}
    \end{minipage}    \hfill
    \begin{minipage}{0.53\textwidth}
    \includegraphics[width=\linewidth]{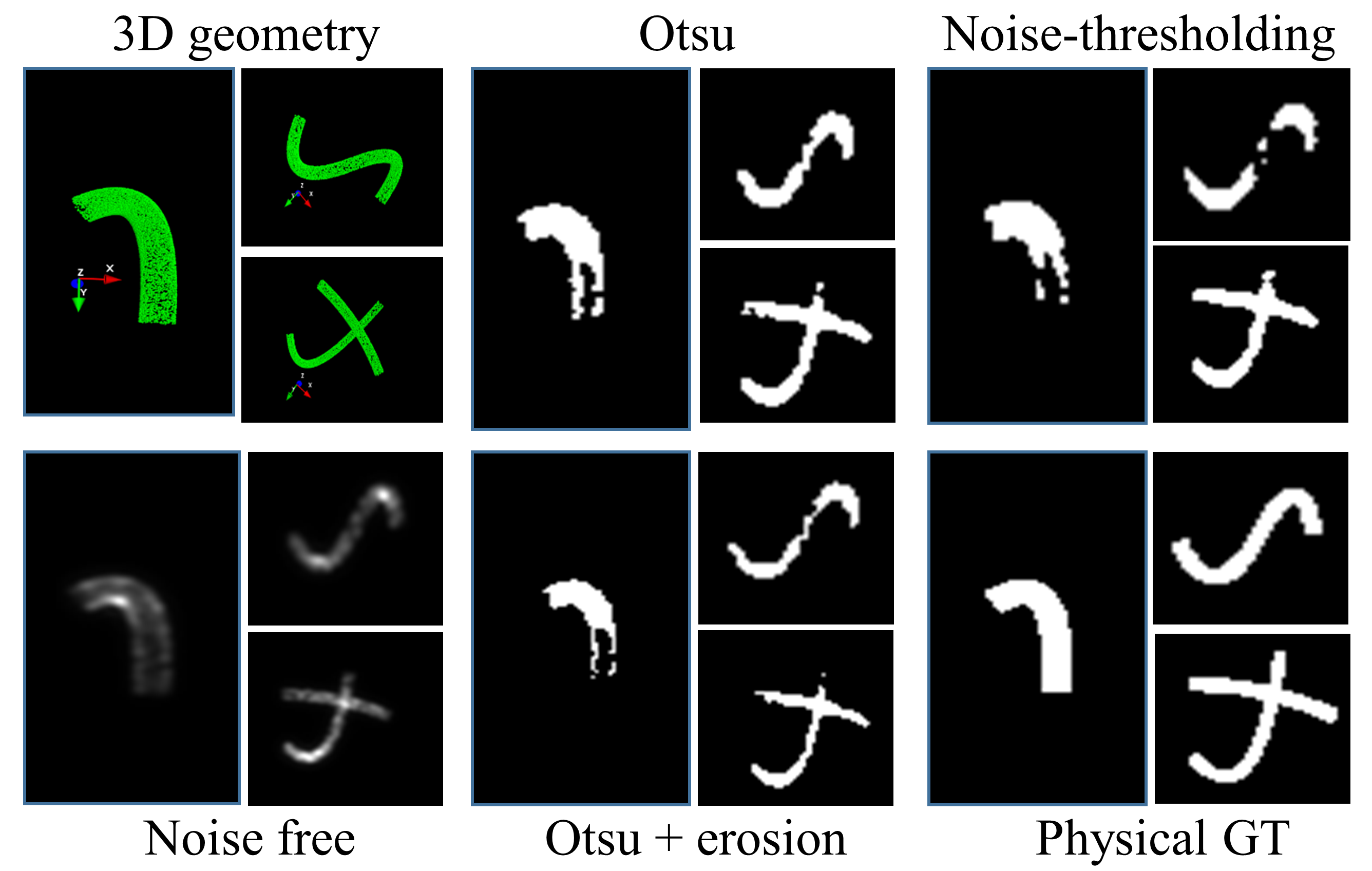}
    \caption{Different techniques of generating ground truth from noise-free simulated images are shown here.}\label{fig:GT_comp}
    \end{minipage}
    \includegraphics[width=\textwidth]{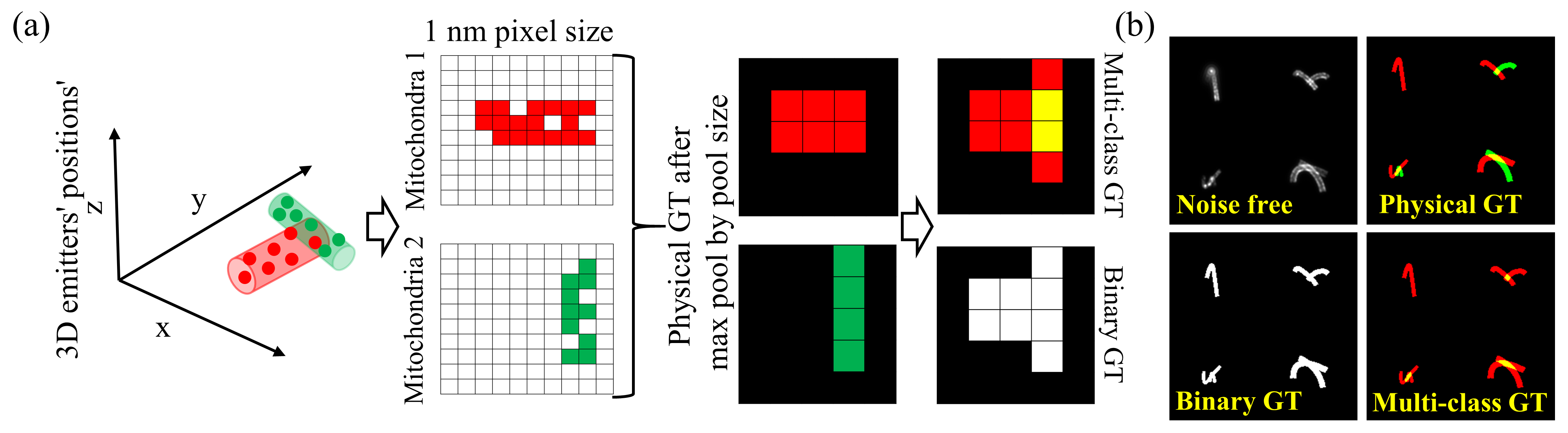}
    \caption{(a) Framework for generating physics based GT. (b) Examples of a simulated sample contains noisy image, noise free, physics based binary GT, and the multi-color GT (red is non overlapping ROI and yellow is overlapping ROI).}
    \label{fig:simulation2}
\end{figure}

In the literature, primarily three types of mitochondrial shapes in 2D microscopy image have been reported, namely dot, rod, and network \cite{leonard2015quantitative}. These shapes are observed due to the different orientation and number of mitochondria present together. Our simulation of single and multiple mitochondria in 3D is able to emulate the reported and observed shapes, as shown in Figure~\ref{fig:dataset}.

\subsection{GT techniques}\label{sec:GT}
Automated generation of segmentation ground truth is the most important task in our proposed method. We considered four approaches, with examples are shown in Figure \ref{fig:GT_comp}. They are Otsu thresholding \cite{xu2011characteristic} on noise-free simulated image, Otsu thresholding followed by erosion filter of half the diameter of PSF in the focal plane, thresholding based on the signal to noise ratio, and physical morphology based ground truth. In noise based thresholding, we compute average noise level from the peak intensity and signal to noise ratio and use that as the threshold.

The generation of physical morphology based GT is illustrated in Figure \ref{fig:simulation2}(a). We first project all the emitter locations into a hypothetical x-y plane with precision $1$ nm (i.e. pixel size of $1$ nm). Then, we binarize this hypothetical image of $1$ nm pixel size by assigning the value $0$ to pixels with no emitters and $1$ otherwise. Then we perform max-pooling with pool size and stride length equal to the actual pixel size (here $80$ nm), followed by binarization, to achieve the ground truth for each mitochondrion. If more than one mitochondrion is present (such as in a network), we process each mitochondrion independently. This helps us in allowing support for multi-class segmentation and classification also. To this end, we assign each mitochondrion one label (or color). For the segmentation problem only, we simply perform boolean AND operation on the corresponding images. To identify the situations of overlapping or network mitochondria, we form a three label ground truth as follows where the labels represent background pixel, pixel corresponding to single mitochondrion, and pixel corresponding to overlapping mitochondria. Some examples are provided in Figure~\ref{fig:simulation2}(b).

There are two main problems with the thresholding based techniques. First, if a mitochondria network is present, the resulting higher intensity at junctions of mitochondria biases the results of thresholding to suppress other mitchondria. Second, the thresholding based methods may either remove out-of-focus information or introduce distortion due to wider PSF in the slightly off-focus region. These problems are absent in the physical GT. At the same time, since 3D-PSF is a convolution operation, its characteristics may be easily learnt and compensated for in the convolutional neural networks. Therefore, it is safe to consider that a perfect segmentation algorithm will produce a binary mask of the entire mitochondrion irrespective its portions being out-of-focus. Based on these arguments as well as some other tests (reported in supplementary), we use physical GT.

\subsection{Segmentation using neural network and its applications}
Although there exist several neural network based segmentation methods in literature, only a small subset of them have been successfully applied to microscopy data. We have noted six of them and used as a baseline for benchmarking our approach and identifying suitable method for our problem. These are (1) a deep CNN~\cite{sadanandan2017automated}, (2) a U-Net~\cite{ronneberger2015u} based approach, (3) a semantic segmentation method~\cite{noh2015learning}, (4) a mask R-CNN method~\cite{he2017mask}, (5) a backpropagation based unsupervised segmentation~\cite{kanezaki2018unsupervised}, and (6) a pixel-to-pixel (Pix2Pix) adversarial network~\cite{tsuda2019cell}. Two separate types of segmentation problems namely binary and multi-class segmentation are benchmarked. The classes corresponds to pixel corresponding to background, a single mitochondrion, or overlapping region of multiple mitochondria. We tested the trained networks on simulated test data and manually annotated experimental data. The details of the manual segmentation approach are provided in the supplementary. The proposed method is compared with competitive baseline thresholding based unsupervised image segmentation methods such as Otsu and adaptive thresholding. Finally, we demonstrate two applications that benefit from segmentation of mitochondria:
\\
\textcolor{white}{space}\textbf{Data analytics of mitochondrial morphologies:} It has been reported that the analytics of dot, rod, and network like mitochondria leads to insights about the health of a cell. The relevant statistics include the number and size of different types of mitochondria in a cell under various cell culture conditions ~\cite{leonard2015quantitative}. The primary challenge of such analysis is accurate segmentation. First, we apply the U-Net trained with the proposed simulation data for segmentation. Next, we employ multi-class segmentation to find overlapping regions and identify network-like mitochondria. Unsupervised classification based on the segmented area of individual mitochondria was used to classify the rod and dot morphologies. This is explained in the supplementary.
\\
\textcolor{white}{space}\textbf{Tracking:} First, the proposed U-Net based
segmentation is used to segment the mitochondria. Next, multi-class segmentation (described above) is used to separate the network mitochondria for individual mitochondria. Then, Kalman filter and Hungarian algorithm \cite{weng2006video} are employed to track individual mitochondria or networks over time~\cite{wojke2017simple}. We propose to use such tracks for spatio-temporal analysis of mitochondria.

\section{Experimental results}
%\subsection{Dataset}
\textbf{Datasets:}
We present two datasets, the simulated dataset for training and the experimental dataset of manually annotated microscopy videos of living cells with fluorescently labelled mitochondria.
% \\
\textbf{\textit{Simulated dataset:}} The simulated dataset contains 9000 simulated noisy images, each of size $256 \times 256$ pixels and emulating the microscope that has been used in the experimental dataset. We explain the formation of an image in this dataset. First, 4 independent simulations are done to form noisy microscopy images of size $128 \times 128$ pixels. Each image simulates either one mitochondrion or a pair of mitochondria (possibly network). Then a montage of them is created to form a noise-free image of size $256 \times 256$ pixels containing multiple mitochondria. Then, the noise is added to the montage to emulate the signal to noise ratio of the experimental dataset.
%\\
\textbf{\textit{Experimental dataset:}} The experimental details of the cells, labeling, sample preparation, microscope and imaging parameters are given in the supplementary. There are 30 videos, each of $\sim$3000 frames and $1024 \times 1024$ pixels. We have manually annotated only the first frame of each video for evaluation in Table \ref{tab:results}.

%\subsection{Baseline and training}
\textbf{Training and baseline:}
The simulation dataset is randomly divided into 3 sets namely training, validation, and testing in the ratio 70:20:10. This approach is used for both binary segmentation and multi-class segmentation. Standard data augmentation such as flip, rotation, cropping, etc. are applied during training. We note that we have not used the conventional image resizing since the physical image pixel size has to be maintained for microscopy data. The validation loss is used as the learning rate reduction and early stopping.
Deep CNN~\cite{sadanandan2017automated} originally utilized an automatic GT generation method using CellProfiler. We instead use our simulation GT for training the network. The original implementation of U-Net~\cite{ronneberger2015u} is used with minor changes to adopt it to our dataset. We experimented with resnet-34, resnet-50, and VGG-16 and found that resnet-34 performs better. We therefore use it as the baseline architecture of U-Net for benchmarking. Semantic segmentation method reported in~\cite{noh2015learning} consists of a set of convolution and deconvolution layers adopted from VGG-16 network. We used the original structure without any major changes. Mask R-CNN~\cite{he2017mask} is also used in its original implementation. The unsupervised method reported in\cite{kanezaki2018unsupervised} is an iterative approach; we used 2000 iterations for segmentation of the mitochondria.
\vspace{-5pt}
\begin{table}[b]
\centering
\caption{Segmentation performance of different methods in simulation dataset (900 test images) and experimental dataset (30 test images from live cell). }
\begin{tabular}{lccccccc}
\hline
\multirow{3}{*}{Type} & \multirow{3}{*}{Method} & \multicolumn{4}{c}{Binary segmentation} & \multicolumn{2}{c}{Multi-class} \\ \cline{3-8}
 &  & \multicolumn{2}{c}{Simulation} & \multicolumn{2}{c}{Live cell} & \multicolumn{2}{c}{Simulation} \\ \cline{3-8}
 &  & mIoU & F1 & mIoU & F1 & mIoU & F1 \\ \hline
\multirow{3}{*}{Unsupervised} &   Otsu~\cite{xu2011characteristic}  & 0.69 & 0.71 & 0.51 & 0.53 & NA & NA \\\cline{2-8}
 & Adaptive thresholding~\cite{shen2018automatic}  & 0.56 & 0.59 & 0.51 & 0.54 & NA & NA \\ \cline{2-8}
 & Backpropagation~\cite{kanezaki2018unsupervised} & 0.68 & 0.71 & 0.58 & 0.62 & 0.61 & 0.64 \\ \hline
Semi-supervised &  ImageJ plugin \cite{ignacio2019}  & 0.76 & 0.79 & 0.72 & 0.79 & NA & NA \\ \hline
\multirow{5}{*}{\begin{tabular}[c]{@{}l@{}}Simulationm-\\ Supervised\end{tabular}}  & Deep CNN~\cite{sadanandan2017automated}  & 0.89 & 0.91 & 0.86 & 0.88 & 0.88 & 0.91  \\ \cline{2-8}
 & U-Net~\cite{ronneberger2015u}  & 0.92 & 0.92 & 0.91& 0.92 & 0.91 & 0.93 \\ \cline{2-8}
   & Semantic segmentation~\cite{noh2015learning}  & 0.92 & 0.94 & 0.81 & 0.89 & 0.90 & 0.92\\ \cline{2-8}
   & Mask R-CNN method~\cite{he2017mask}  & 0.89 & 0.91 & 0.79 & 0.88 & NA & NA \\ \cline{2-8}
  & Pix2Pix GAN~\cite{tsuda2019cell}  & 0.84 & 0.86 & 0.79 & 0.82 & 0.89 & 0.91 \\ \cline{2-8} \hline
\end{tabular}
\label{tab:results}
\end{table}

\begin{figure}[h]
\centering
\includegraphics[width=\linewidth]{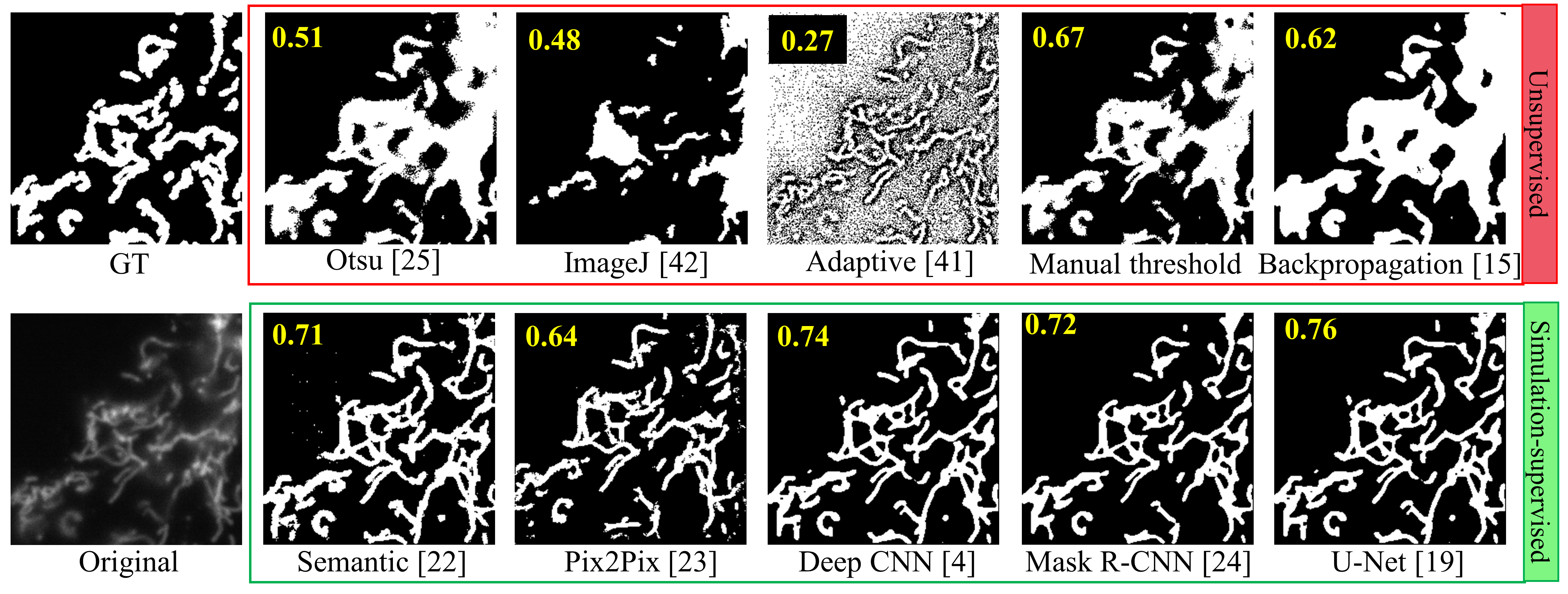}\vspace{-2mm}
\caption{Examples of a segmentation using different methods. The numbers denote mIoU.}
\label{fig:examples}
\end{figure}

%\subsection{Binary and multi-class segmentation results}
\textbf{Binary and multi-class segmentation results:}
Here, we present the results of binary segmentation and multi-class segmentation. We divide the analysis in three sets. The first set corresponds to baseline {unsupervised} methods such as Otsu, adaptive threshold, and backpropagation based unsupervised segmentation. The second set corresponds to {segmentation by expert using ImageJ plugin}. The third set corresponds to supervised learning methods using simulation-supervised training samples. We evaluate the methods using mean intersection-over-union (mIoU) and mean F1 score over background and mitochondria for binary segmentation and background, non overlapping, overlapping pixels in multi-class segmentation. The results are summarized in Table~\ref{tab:results}. It is noted that the simulation-supervised U-Net and deep CNN outperform not just in simulated test data, but also for the experimental data without retraining. Figure~\ref{fig:examples} shows visual comparison of different segmentation methods in a complex area of a living cell.

\begin{figure}[h]
\includegraphics[width=\linewidth]{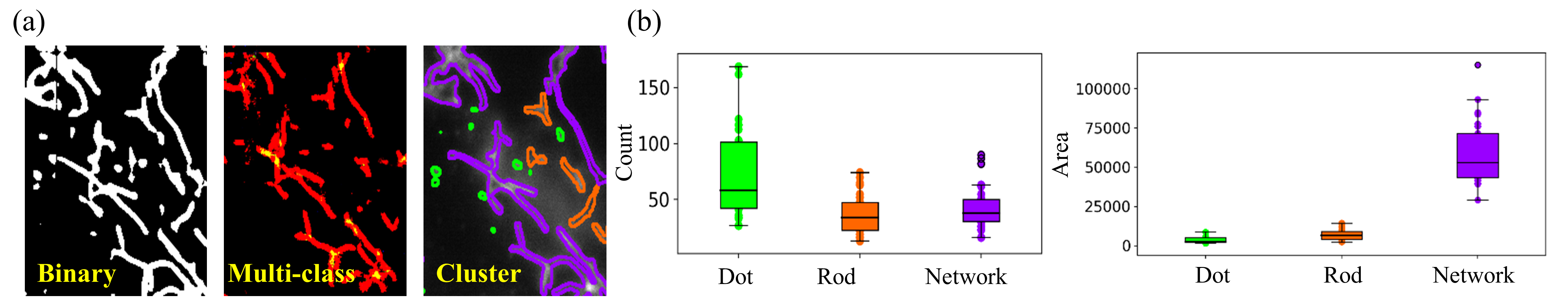}\vspace{-2mm}
\caption{Our approach enables automated statistical analysis of different mitochondrial morphologies. (a) Binary, multi-class segmentation, and clusters of different mitochondrial morphologies (green-dot, orange-rod, purple-network. (b) Statistics of counts (left) and area (right) of each morphology.}
\label{fig:analysis}
\end{figure}
\begin{figure}[h]
\includegraphics[width=\linewidth]{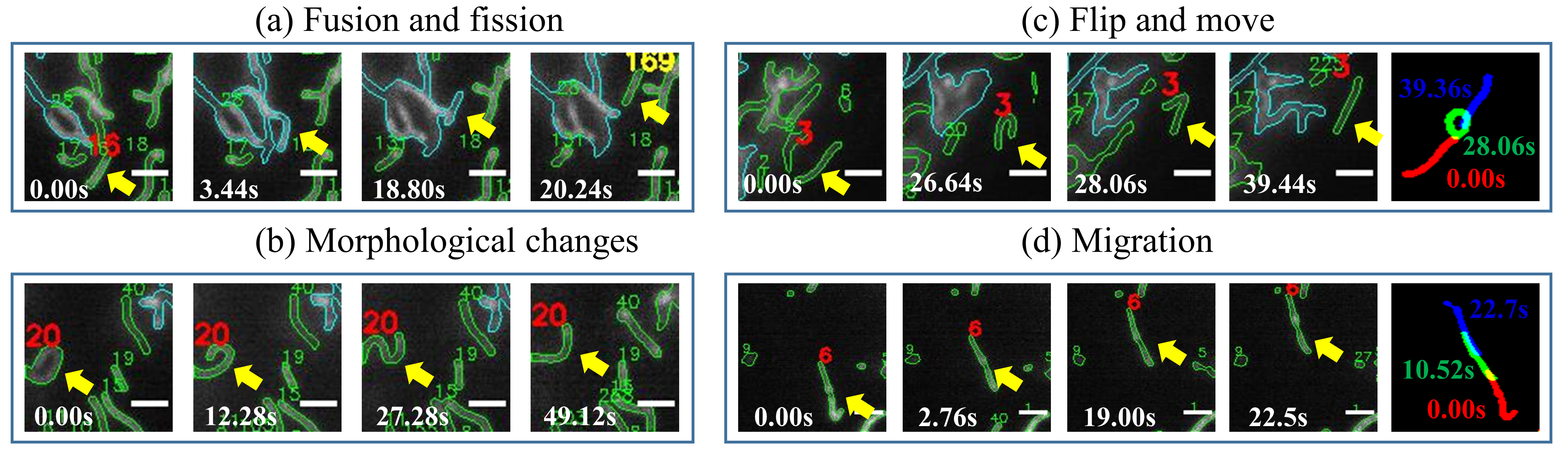}\vspace{-2mm}
\caption{Examples of events detected by mitochondrial segmentation and tracking (scale bar: 2 $\mu m$).}
\label{fig:examples2}
\end{figure}

%\subsection{Mitochondria analysis and tracking}
\textbf{Mitochondrial morphology analytics and tracking:} We discuss the statistical analysis of the number and area of mitochondria of the different different morphologies (dot, rod, and network). The statistics are extracted from the segmentation obtained using simulation-supervised U-Net. Results shown in Figure~\ref{fig:analysis} indicate the possibility of performing large scale biological studies with cells subjected to different pathological conditions and therapeutic treatment such as in cardiac, and kidney diseases \cite{signorile2019prohibitins}.

We also present different events extracted by tracking the segmented mitochondria. We note different motion patterns that may be interest to biologists. Extraction of such movement patterns is possible due to the accurate segmentation and tracking. Figure~\ref{fig:examples2}(a) shows that a mitochondrion moves towards mitochondria network, merges with it (fusion), separates from it after a few seconds (fission), and the moves away. Figure~\ref{fig:examples2}(b) shows a typical morphological change of a mitochondrion (circular blob to curve). In, Figure~\ref{fig:examples2}(c), a mitochondrion flips and moves. Figure~\ref{fig:examples2}(d) shows a mitochondrion migrating from one place to another. The mitochondrion in Figure~\ref{fig:examples2}(d) was segmented by the expert as two mitochondria (see supplementary) until the expert observed that the mitochondrion moves as a single entity. On the other hand, simulation-supervision segmented it as a single mitochondrion, compensating automatically for the out-of-focus region in the middle of it.

\section{Discussion and conclusions}
We demonstrate that semi-supervised segmentation is not just feasible for the challenging task of mitochondrial segmentation, it opens up several opportunities for performing advanced computer vision tasks towards generating valuable knowledge in biological studies. We bring physics and AI to a nexus where AI can create a significant impact with a little help from physics based modeling. It eradicates the problems of subjectivity, variability, and inaccuracy associated with manual segmentation. At the same time, it enables automatic creation of large datasets custom-designed for the specific application at hand. We note that designing the simulator to emulate the system as accurately as possible is imperative for the concept of simulation-supervised to be useful. We show two examples of this in the supplementary.
Here, we have shown feasibility of segmentation and then tracking of mitochondria. But, the concept of simulation-supervision is scalable to generate supervised datasets for dynamic behaviours such as seen in Figure \ref{fig:examples2}, or to perform AI tasks on other extremely challenging and biologically significant structures such as the endoplasmic reticulum. Therefore, we see simulation-supervised learning approaches as the ingress of CV into the next generation microscopy and other GT-hard scientific problems.

The first wave of CV converted commercial cameras from aesthetic and memory keeping devices to smart tools in our day-to-day life. Through our work, we aim to create a second wave of CV that converts microscopes from visualization devices to knowledge discovery tools for biology. We envisage multi-dimensional impact of this work, some of which we list below
\begin{itemize}
    \item High content microscopy analytics: Demonstrating that simulation supervised can be the solution to perform AI is a big leap towards automated high content microscopy data analysis for drug trials. High throughput imaging is the new paradigm, but analyzing these big data manually is stopping in realization of its potential. Simulation-supervised learning tools will be the solution
    \item Rare event detection: Microscopy data is analogous to a sea of information in which the critical events are rare in both spatial and temporal sense. So, after acquiring microscopy time lapse videos, biologists spend weeks to months looking for that signature of a sparse event which might have triggered death or recovery. The hypotheses for mechanisms, morphology, and dynamics may have been reported, but finding evidence is often a tedious task, which AI can easily undertake with good simulation-supervised learning paradigms.
    \item Analytics from single entity to large sample pools: Imagine how the biologist will be empowered if every mitochondrion in a cell can be tracked, its events and interactions can be mapped, and it is straight forward to derive exploitable analytics at all the relevant scales, i.e. from mitochondrian-scale to cell-scale, cell-colony scale, and then large population scale. This aspect is generalizable to a huge variety of clinically relevant sub-cellular entities.
    \item Beyond microscopy: Prognosis of a disease or syndrome is critical to design therapy. Generating ground truth may require several monitoring patients for years. Such approach is indispensable, but simulation-supervised modeling of prognosis based on known hypotheses can be quite useful before big medically qualified datasets are available. Similarly for prediction problems, rare phenomena detection, and other AI problems in climate, epidemiology, etc., simulation-supervised learning can show the pointer.
\end{itemize}

\bibliographystyle{unsrtnat}
\bibliography{mybib}   % name your BibTeX data base

\end{document}